\documentclass[preprint,aps]{revtex4}

\usepackage{graphicx}

\begin{document}

\title{Observation of Flat Band, Dirac Nodal Lines and Topological Surface States in Kagome Superconductor CsTi$_3$Bi$_5$}
\author{Jiangang Yang$^{1,2,\sharp}$, Yuyang Xie$^{1,2,\sharp}$, Zhen Zhao$^{1,2,\sharp}$, Xinwei Yi$^{2,3,\sharp}$, Taimin Miao$^{1,2}$, Hailan Luo$^{1,2}$, Hao Chen$^{1,2}$, Bo Liang$^{1,2}$, Wenpei Zhu$^{1,2}$, Yuhan Ye$^{1,2}$, Jing-Yang You$^{4}$, Bo Gu$^{2,3}$, Shenjin Zhang$^{5}$, Fengfeng Zhang$^{5}$, Feng Yang$^{5}$, Zhimin Wang$^{5}$, Qinjun Peng$^{5}$, Hanqing Mao$^{1,2,6}$, Guodong Liu$^{1,2,6}$, Zuyan Xu$^{5}$, Hui Chen$^{1,2,3}$, Haitao Yang$^{1,2,3}$, Gang Su$^{2,3*}$, Hongjun Gao$^{1,2,3,*}$, Lin Zhao$^{1,2,6,*}$ and X. J. Zhou$^{1,2,6,*}$
}

\affiliation{
\\$^{1}$Beijing National Laboratory for Condensed Matter Physics, Institute of Physics, Chinese Academy of Sciences, Beijing 100190, China.
\\$^{2}$University of Chinese Academy of Sciences, Beijing 100049, China.
\\$^{3}$CAS Center for Excellence in Topological Quantum Computation, University of Chinese Academy of
Sciences, Beijing 100190, China
\\$^{4}$Department of Physics, Faculty of Science, National University of Singapore, Singapore 117551, Singapore.
\\$^{5}$Technical Institute of Physics and Chemistry, Chinese Academy of Sciences, Beijing 100190, China.
\\$^{6}$Songshan Lake Materials Laboratory, Dongguan, Guangdong 523808, China.
}
\date{December 8, 2022}
\pacs{}

\begin{abstract}

A kagome lattice of 3d transition metals hosts flat bands, Dirac fermions and saddle points. It provides a versatile platform for achieving topological superconductivity, anomalous Hall effect, unconventional density wave order and quantum spin liquid when the strong correlation,  spin-orbit coupling or magnetic order are involved in such a lattice.
Here, using laser-based angle-resolved photoemission spectroscopy in combination with density functional theory calculations, we investigate the electronic structure of the newly discovered kagome superconductor CsTi$_3$Bi$_5$, which is isostructural to the AV$_3$Sb$_5$ (A=K,\,Rb or Cs) kagome superconductors and possesses a perfect two-dimensional kagome network of Titanium.
We directly observed a strikingly flat band derived from the local destructive interferences of Bloch wave functions within the kagome lattices.
We also identify the type-II Dirac nodal loops around the Brillouin zone center, the type-III Dirac nodal loops around the zone corners and type-III Dirac nodal lines along the k$_z$ direction. In addition, around the Brillouin zone center, Z2 nontrivial topological surface states are also observed which is formed from the band inversion due to strong spin orbital coupling. The simultaneous  existence of such multi-sets of nontrivial band structures in one kagome superconductor not only provides good opportunities to study related physics in the kagome lattice  but also makes CsTi$_3$Bi$_5$ an ideal system to realize noval quantum phenomena by manipulating its chemical potential with chemical doping or pressure.

\end{abstract}

\maketitle

\newpage

Quantum materials with layered kagome structures have drawn enormous attentions because such a two-dimensional (2D) network of corner-sharing triangle lattice can give rise to many exotic quantum phenomena, such as spin liquid phases\cite{SSachdev1992,Patrick2008,LBalents2010,SWhite2011SYan}, topological insulator and topological superconductor\cite{MFranz2009HMGuo,RThomale2012MLKiesel,RThomale2013MLKiesel}, fractional quantum Hall effect\cite{XGWen2011ETang}, quantum anomalous Hall effect\cite{CFelser2014JKubler,SCZhang2015GXu} and unconventional density wave orders\cite{JXLi2012SLYu,20120904_QHWang2013WSWang,SDWilson2020BROrtiz,20210316_HJGao2021HChen,MZHasan2021YXJiang,PDai2022XTeng}. All these exotic quantum phenomena are thought to originate from the unique electronic structure of the kagome lattice including flat bands, Dirac cones and saddle points  when the spin orbital coupling, magnetic ordering or strong correlation are taken into consideration. Nevertheless, the definitive identification of such unique electronic structures in the kagome materials are still scarce and the underlying mechanism to induce those exotic quantum phenomena from such electronic structures remains illusive. For example, the kagome superconductors AV$_3$Sb$_5$ (A=K, Rb or Cs) \cite{SDWilson2019BROrtiz1}, which have been the focus of recent extensive research, exihibit  anomalous Hall effect\cite{20200305_MNAli2020SYYang}, unconventional charge density wave (CDW)\cite{MZHasan2021YXJiang,20210126_HCLei2021QWYin,20210310_BHYan2021HXTan,20210312_JPHu2021XLFeng,20210328_SAYang2021JZZhao,XHChen2021ZWLiang,YGYao2021ZWWang,IZeljkovic2021HZhao}, pairing density wave\cite{20210316_HJGao2021HChen} and possible unconventional superconductivity and nematic phase\cite{SDWilson2020BROrtiz,20200305_MNAli2020SYYang,XHChen2022LPNie,IZeljkovic2021HZhao}. However, the nature and origin of these noval physical properties are still in hot debates. Even for the clear identification of the flat band, it still needs further investigations. It is significant to establish a relationship between the unique electronic structures of the kogome lattice and its novel quantum phenomena.

CsTi$_3$Bi$_5$ is a newly discovered kagome superconductor which is isostructural to the  AV$_3$Sb$_5$ superconductors (Fig. 1a)\cite{SGang2022XWYi}. The titanium atoms form a kagome network with the bismuth atoms lying in the hexagons and above and below the triangles (Figs. 1a and 1b). Magnetic susceptibility and electrical resistivity measurements of CsTi$_3$Bi$_5$  indicate that there is no phase transition observed down to the superconducting  transition at 4.8\,K\cite{HaitaoYang2022HJGao}. This is different from CsV$_3$Sb$_5$ that exhibits a CDW transition around 94\,K\cite{SDWilson2019BROrtiz1}. The similar crystal structure but the absence of the CDW order in CsTi$_3$Bi$_5$ provide a good opportunity to study the intrinsic electronic structure of the kagome lattice with reference to CsV$_3$Sb$_5$ and understand the origin of various quantum phenomena in kagome materials.

In this paper, we investigate the electronic structure of the newly discovered kagome superconductor CsTi$_3$Bi$_5$. By using high resolution laser-based angle-resolved photoemission spectroscopy (ARPES), in combination with the band structure calculations, we have directly observed the characteristic electronic features of the kagome lattice. We directly observed the flat band derived from the destructive interferences of the Bloch wave functions within the kagome lattices. We also identify the Dirac nodal loops and nodal lines in three dimensional momentum space. The Z2 nontrivial topological surface states are also observed. Such coexistence  of multiple nontrivial band structures in one kagome superconductor provides a new platform to study the rich physics in the kagome lattice.

CsTi$_3$Bi$_5$ single crystals were grown using a self flux method\cite{HaitaoYang2022HJGao}.  Typical CsTi$_3$Bi$_5$ crystals with a lateral size of $\sim$3\,mm and regular hexagonal morphology were obtained. High resolution angle-resolved photoemission measurements were performed using a lab-based ARPES system equipped with the 6.994 eV vacuum-ultra-violet (VUV) laser and  a hemispherical electron energy analyzer DA30L (Scienta-Omicron)\cite{XJZhou2008GDLiu,XJZhou2018}. The laser spot is focused to around 10\,um on the sample in order to minimize  the influence of sample inhomogeneity. The light polarization can be varied to get linear polarization along different directions. In the LV (LH) polarization the electric vector of the laser light is perpendicular (parallel) to the plane formed by the light path and the analyzer lens axis. The energy resolution was set at 1 meV and the angular resolution was 0.3 degree corresponding to 0.004 ${\AA}^{-1}$ momentum resolution at the photon energy of 6.994 eV. All the samples were cleaved {\it in situ} at a low temperature of 20\,K and measured in ultrahigh vacuum with a base pressure better than 5 x 10$^{-11}$ mbar. The Fermi level is referenced by measuring on clean polycrystalline gold that is electrically connected to the sample.

Density functional theory (DFT) calculations with projector augmented-wave pseudopotential method\cite{DFT1} are implemented through Vienna ab initio simulation package (VASP)\cite{DFT2}. The exchange correlation functional is treated by Perdew-Burke-Ernzerh (PBE) of parameterization of generalized gradient approximation (GGA)\cite{DFT3}. The convergence criterion of atomic forces in structural optimization with VASP is less than 1\,meV/$\AA$  total energy convergence threshold of all processes is 10$^{-6}$ eV/atom. The cutoff energy of the plane-wave is set as 520 eV. The $\Gamma$ centered 20$\times$20$\times$12 Monkhorst-Pack k-point grid is used in the self-consistent cycle. Wannier90 package\cite{DFT4} is used to fit Wannier functions and construct tight-binding models, and WannierTools\cite{DFT5} package is used to calculate the surface spectral functions by using the surface Green's function method. Calculations of structures' parity are performed through a combination of the irvsp program\cite{DFT6} and VASP.


 Figure 1d shows the Fermi surface mapping of CsTi$_3$Bi$_5$ measured at 20\,K. The entire first BZ is covered by our laser ARPES measurements. Five Fermi surface sheets are clearly observed, as quantitatively shown in Fig. 1e. The Fermi surface consists of three electron-like  Fermi surface sheets around  $\bar{\Gamma}$ ($\alpha$, $\beta$ and $\gamma_1$ in Fig. 1e), an electron-like triangular Fermi pocket around $\bar{K}$ ($\gamma_2$ in Fig. 1e) and a small hole-like Fermi pocket around $\bar{M}$ ($\delta$ in Fig. 1e).

 In order to understand the measured electronic structure, we carried out detailed band structure calculations. Figs. 1f and 1g show the calculated band structures of CsTi$_3$Bi$_5$ without considering the spin orbit couping (SOC) (Fig. 1f) and considering SOC (Fig. 1g). These calculations project the band structures onto different Ti\,3d orbitals along high symmetry directions in the BZ. The low energy bands are mainly from the 3d orbitals of titanium. The characteristic electronic features of a kagome lattice, including the flat band, two saddle points at $\bar{M}$ and a Dirac point at $\bar{K}$, can be clearly observed as marked in Figs. 1f and 1g. These features are mainly from the Ti\,3d$_{{x^2-y^2}/xy}$ orbitals (red lines in Figs. 1f and 1g) except that the saddle point VHS1 is from Ti\,3d$_{z^2}$ (green line in Figs. 1f and 1g). The consideration of SOC shows little effect on the flat band and the saddle points but opens a gap at the Dirac points (Fig. 1g).

 The calculated band structures of CsTi$_3$Bi$_5$ (Figs. 1f and 1g)  are very similar to that of CsV$_3$Sb$_5$ where the kagome lattice related electronic features are mainly from the V\,3d orbitals\cite{SDWilson2020BROrtiz,DWilson2021BROrtiz}. The main difference is the band position with respect to the Fermi level. In CsTi$_3$Bi$_5$ the kagome lattice related bands are shifted upwards by $\sim$1\,eV when compared with those in CsV$_3$Sb$_5$. This is because CsTi$_3$Bi$_5$ has one electron less per Ti per unit cell than that of CsV$_3$Sb$_5$ when Ti is replaced by V. As a result, although the Ti\,3d orbitals still dominate the density of states (DOS) around the Fermi level E$_F$, the two van Hove singularities (VHS) in CsTi$_3$Bi$_5$ are above the Fermi level whereas they are close or below the Fermi level in CsV$_3$Sb$_5$\cite{XJZhou2022HLLuo,RComin2022MKang}. This provides a possible explanation of the absence of the CDW order in CsTi$_3$Bi$_5$. The upward band shift also moves the flat band close to Fermi level in CsTi$_3$Bi$_5$.


Figure 1h shows the calculated Fermi surface in the three dimensional Brillouin zone. The Fermi surface consists of five  sheets which are quite two dimensional. This is expected due to the strong in-plane bonding and weak interlayer coupling in CsTi$_3$Bi$_5$ which is similar to that in CsV$_3$Sb$_5$. The calculated Fermi surface at k$_z$=0 is shown in Fig. 1i. To make a better comparison between the measured Fermi surface and band structures with the band structure calculations, we find that the Fermi level of the calculated band structures needs to be shifted downwards by $\sim$90\,meV, as shown in Figs. 1f and 1g. The calculated Fermi surface (Figs. 1h and 1i) shows an excellent agreement with the measured results in Figs. 1d and 1e.


\textbf{Nontrivial Flat Band}

For the genuine kagome lattice, it shows a perfect flat band across the entire Brillouin zone, as schematically shown in Fig. 2g. In real kagome materials like CsTi$_3$Bi$_5$, as shown in Figs. 1f and 1g, the flat band (FB) is nearly dispersionless along $\bar{K}$-$\bar{M}$-$\bar{K}$ but becomes dispersive along $\bar{\Gamma}$-$\bar{M}$ and $\bar{\Gamma}$-$\bar{K}$ directions. This is because, in real kagome materials, the flat band dispersion
can be modified by additional factors besides the spin orbit coupling, such as the in-plane next-nearest-neighbor hopping, the interlayer coupling or the multiple orbital degrees of freedom. So far there have been some ARPES studies reporting the  observation of the kagome lattice-derived flat band in GdV$_6$Sn$_6$\cite{JFHe2021STPeng}, YMn$_6$Sn$_6$\cite{SCWang2021MLi}, CoSn\cite{RComin2020MGKang,SCWang2020ZHLiu1,CGZeng2022HHuang}, Fe$_3$Sn$_2$\cite{ZZhang2018ZLin} and FeSn\cite{RComin2020MGKang1}. However, there is little clear evidence reported about the kagome-derived flat band in the 135 family represented by AV$_3$Sb$_5$(A=K, Rb or Cs).

Figure 2a and 2b show the band structures of CsTi$_3$Bi$_5$ measured at 20\,K along the $\bar{\Gamma}$-$\bar{M}$-$\bar{K}$-$\bar{\Gamma}$ high symmetry directions  under the LV (Fig. 2a) and LH (Fig. 2b) light polarizations. In order to resolve all the band structures more clearly, the corresponding second derivative image is shown in Fig. 2c. There seems to be a dispersionless band across the entire Brillouin zone at the binding energy of $\sim$220\,meV. A careful analysis indicated that this band consists of two different parts. The first part is marked by the red dashed line in Fig. 2c while the rest of the band represent the second part. As compared with the band structure calculations in Figs. 2d and 2e, the first part shows a good agreement with the flat band from the band structure calculations. This indicates that it is kagome lattice derived flat band. This band can be attributed to the local destructive interferences of the Bloch wave functions within the kagome lattices (Fig. 2f). The second part of the dispersionless band at $\sim$220\,meV is not expected from the band structure calculations (Figs. 2d and 2e). A careful inspection indicates that there is an additional spectral weight buildup in the binding energy range of 220$\sim$500\,meV. The two energies happen to coincide with the top and bottom energy positions of the first part flat band. This indicates that the extra spectral weight buildup is closely related to the first part flat band. The second part flat band at $\sim$220\,meV actually represents a spectral weight cut-off at this energy (see Fig. S1 in Supplementary Materials for details).

\textbf{Dirac Nodal Lines}

In some quantum materials, the bands can cross at a discrete point in the momentum space, forming Dirac point with spin degeneracy or Weyl point with spin polarization. The Dirac points can also form nodal lines and nodal loops in three dimensional momentum space\cite{YLChen2020CFZhang}. The Dirac points can be categorized into three types according to the slopes of the involved bands\cite{BABernevig2015AASoluyanov,XJZhou2022CYSong}.  The materials, which have the electronic structure with the type-II (two dispersion branches exhibit the same sign of slope) or type-III (one of two branches is dispersionless) Dirac points, may host exotic properties, e.g., the chiral anomaly\cite{BABernevig2015AASoluyanov} and Klein tunneling\cite{CWJBeenakker2015TEOBrien}. However, there have been few established cases of the type-II and type-III Dirac point realization in real materials, not to mention their  simultaneous observation in one material.

Figure 3a and 3b show our identification of two sets of Dirac nodal loops and one set of Dirac lines in CsTi$_3$Bi$_5$.  Fig. 3a  shows the calculated band structure along the high-symmetry directions without considering spin-orbit coupling. We can find two groups of linear dispersion crossings in a covered energy region around E$_F$, marked as NL1 and NL2 in Fig. 3a. Our DFT calculations reveal that these Dirac nodes are not isolated, but form multiple nodal loops in k$_z$=0 and k$_z$=$\pi$/c planes as seen in Fig 3b. The NL1 type-II nodal loops form in-plane hexagons centered on $\Gamma$ and A while the NL2 nodal loops form in-plane triangles centered on all the K and H points. These nodal loops are protected by the M$_z$ mirror symmetry. Detailed band analysis shows that the type-II NL1 of k$_z$=0 and k$_z$=$\pi$/c planes are not connected along the k$_z$ direction due to the absence of mirror symmetry between 0$<$k$_z$$<$$\pi$/c. However, for the type-III NL2 in the k$_z$=0 and k$_z$=$\pi$/c planes, we find another set of nodal lines in the $\Gamma$-K-H-A plane that links them. These nodal lines are type-III and protected by the M$_x$ mirror symmetry. Due to the six-fold rotational symmetry, there are six nodal lines and NL2 loops that are symmetrically distributed near K and L points. Slices at different k$_z$ positions have similar band structures, which makes these nodal loops in different slices  still possible to be captured experimentally in spite of the opening of small gaps. Moreover, after considering SOC, these nodal loops will further open gaps but the gap size remains small ($<$50meV).

Figure 3c to 3e show the measured band structures along $\bar{\Gamma}$-$\bar{M}$, $\bar{M}$-$\bar{K}$ and $\bar{K}$-$\bar{\Gamma}$ high-symmetry directions, respectively. For comparison, the corresponding calculated band structures with SOC are presented in Figs. 3f to 3h. The calculated bands agree very well with the experimental results. The NL1 point is formed by the crossing of the $\beta$ and $\gamma_1$ bands, as shown by the blue and red dashed lines in Figs. 3c and 3e. These two bands ($\beta$ and $\gamma_1$) share the same sign of slope along both $\bar{\Gamma}$-$\bar{M}$ (Fig. 3c) and $\bar{\Gamma}$-$\bar{K}$ (Fig. 3e) directions, forming a type-II Dirac nodal loop. The NL2 point is formed by the crossing of the $\gamma_2$ band and the kagome flat band, as shown by the blue and red dashed lines in Figs. 3d and 3e. Since the kagome flat band is nearly dispersionless, the NL2 Dirac nodal loop can be categorized into type-III.

\textbf{Nontrivial Topological Surface States}

The spin-orbit coupling is stronger in CsTi$_3$Bi$_5$ than that in CsV$_3$Sb$_5$ because of the heavy element Bi. We also note that the calculated energy bands give rise to a strong topological Z2 index  since the PT symmetry is conserved in CsTi$_3$Bi$_5$ (see Fig. S3 in Supplementary Materials)\cite{HaitaoYang2022HJGao, SGang2022XWYi}. This will result in possible topologically nontrivial surface states. Fig. 4a shows the band structure measured around $\bar{\Gamma}$ along the $\bar{M}$-$\bar{\Gamma}$-$\bar{M}$ direction under the LV light polarization. The corresponding
second derivative image is shown in Fig. 4b. For comparison, Figs. 4c and 4d show the calculated band structures without and with SOC, respectively, along the same momentum cut. All the observed bands in Fig. 4b can be well assigned by comparing with the calculated bands (as shown by coloured lines in Figs. 4b and 4d) except for one band that is marked as TSS in Fig. 4b. In order to understand its origin, we analyzed the energy bands in details. We find that CsTi$_3$Bi$_5$ has symmetry-protected band degeneracy along the $\Gamma$-A path between the $\gamma$ and $\beta$ bands, as well as between the $\beta$ and $\alpha$ bands, giving rise to multiple topological Dirac semimetal states (see Fig. S3 in Supplementary Materials). This type of topological Dirac semimetal states also appear in AV$_3$Sb$_5$ near the Fermi level along the $\Gamma$-A path. However, in CsTi$_3$Bi$_5$, continuous band gaps throughout the whole Brillouin zone exist between the $\epsilon$ and $\delta$ bands, as well as between the $\delta$ and $\gamma$ bands. Combining the time-reversal and inversion symmetries in CsTi$_3$Bi$_5$, we obtain nontrivial Z2 topological invariant of $\epsilon$ and $\delta$ bands by calculating the parity of the wavefunctions at all time-reversal invariant momenta (TRIM)\cite{Kane2007LFu}, as seen in Fig. S3 in Supplementary Materials. Moreover, band gaps and band inversions due to the strong SOC of the system can induce additional Dirac topological surface states (TSS) crossing at the TRIM $\Gamma$ point as seen in the surface spectral function of Figs. 4e and 4f. Comparing with the bands in Fig. 4d, we can identify that the topological surface states are located between the $\epsilon$ and $\delta$ bands, which indicates that they are topologically protected by the nontrivial Z2 invariant. This TSS band gives a good match with the unassigned band in Fig. 4b. Therefore, we provide a definitive spectroscopic evidence that nontrivial topological surface states exist in the kagome superconductor CsTi$_3$Bi$_5$.


In summary, by using high resolution laser based ARPES in combination with the DFT band structure calculations, we investigate the electronic structure of the newly discovered kagome superconductor CsTi$_3$Bi$_5$. The observed Fermi surface and band structures show excellent agreement with the band structure calculations.  We have identified multi-sets of nontrivial band structures in CsTi$_3$Bi$_5$ including the kagome lattice derived flat band, type-II and type-III Dirac nodal loops and nodal lines, as well as Z2 nontrivial topological surface states. Such coexistence of nontrivial band structures in one kagome superconductor provides a new platform to understand the physics and explore for new phenomena and exotic properties in the kagome materials.

$^{\sharp}$These people contribute equally to the present work.

$^{*}$Corresponding authors: XJZhou@iphy.ac.cn, LZhao@iphy.ac.cn, hjgao@iphy.ac.cn, gsu@ucas.ac.cn


%

\vspace{3mm}

\noindent {\bf Acknowledgements}\\
This work is supported by the National Natural Science Foundation of China (Grant No. 11888101, 11922414 and 11974404), the National Key Research and Development Program of China (Grant No. 2021YFA1401800, 2017YFA0302900, 2018YFA0704200, 2018YFA0305600,2019YFA0308000 and 2022YFA1604200), the Strategic Priority Research Program (B) of the Chinese Academy of Sciences (Grant No. XDB25000000 and XDB33000000) , the Youth Innovation Promotion Association of CAS (Grant No. Y2021006) and Synergetic Extreme Condition User Facility (SECUF).

\vspace{3mm}

\noindent {\bf Author Contributions}\\
 J.G.Y., Y.Y.X., Z.Z. and X.W.Y. contribute equally to this work.  X.J.Z., L.Z., J.G.Y. and Y.Y.X. proposed and designed the research. Z.Z., H.C., H.T.Y., H.T.Y. and H.J.G. contributed to single crystal growth. X.W.Y., J.Y.Y., B.G. and G.S. contribute to the DFT band calculations.  T.M.M., H.L.L., H.C., B.L., W.P.Z., S.J.Z, F.F.Z., T.Y., Z.M.W., Q.J.P., H.Q.M., G.D.L., L.Z., Z.Y.X. and X.J.Z. contributed to the development and maintenance of Laser-ARPES system. J.G.Y. and Y.Y.X. carried out the ARPES experiment. J.G.Y., L.Z. and X.J.Z. analyzed the data.  J.G.Y., L.Z. and X.J.Z. wrote the paper with H.T.Y., H.J.G.and G.S.. All authors participated in the discussion and comment on the paper.

\vspace{3mm}

\noindent {\bf\large Additional information}\\
\noindent{\bf Competing financial interests:} The authors declare no competing financial interests.

\newpage

\begin{figure*}[tbp]
\begin{center}
\includegraphics[width=1.0\columnwidth,angle=0]{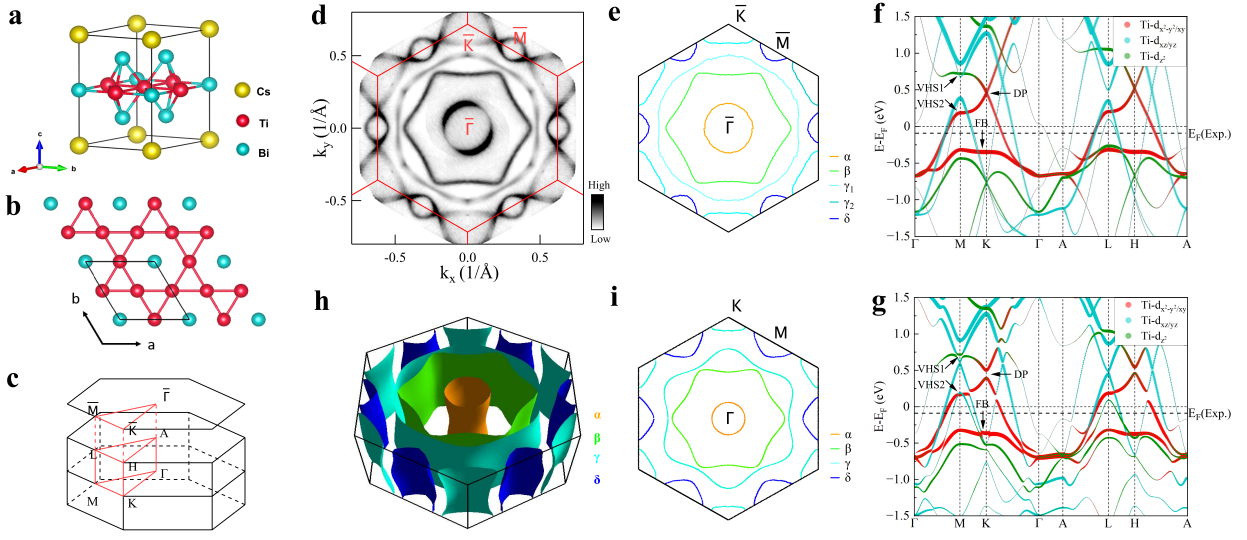}
\end{center}
\caption{\textbf{Fermi surface and calculated band structures of CsTi$_3$Bi$_5$.} \textbf{a} Schematic pristine crystal structure of CsTi$_3$Bi$_5$. \textbf{b} Top view of the crystal structure with a two-dimensional kagome lattice of Titannium. \textbf{c} Three-dimensional Brillouin zone with high-symmetry points and high-symmetry momentum lines marked. \textbf{d} Fermi surface mapping of CsTi$_3$Bi$_5$ measured at a temperture of 20\,K. It is obtained by integrating the spectral intensity within 10\,meV with respect to the Fermi level and symmetrized assuming six-fold symmetry. Five Fermi surface sheets are clearly observed and quantitatively shown in \textbf{e}. Three Fermi surface sheets are around the Brillouin zone center $\Gamma$ marked as $\alpha$ (orange line), $\beta$ (green line) and $\gamma_1$ (light blue line). One Fermi surface is around the K point marked as $\gamma_2$ (blue line) and one is around the M point marked as $\delta$ (dark blue line).
\textbf{f}  Calculated band structure along high-symmetry directions without considering SOC. Different colors represent different orbital components of Ti$_{3d}$. \textbf{g} Same as \textbf{f} but considering SOC. The flat band (FB), two saddle points (VHS1 and VHS2) and a Dirac point (DP) are marked by arrows. To make a better comparison with measured results, the Fermi level is shifted downwards by 90\,meV, as shown by the dashed lines in \textbf{f} and \textbf{g}.	\textbf{h} Calculated three dimensional Fermi surface based on the first principle DFT calculations. The Fermi surface sheets are quite two dimensional. The calculated Fermi surface at k$_z$=0 is shown in \textbf{i}. The measured Fermi surface (\textbf{d}) shows an excellent agreement with the calculated one (\textbf{i}).
}
\end{figure*}

\begin{figure*}[tbp]
\begin{center}
\includegraphics[width=1.0\columnwidth,angle=0]{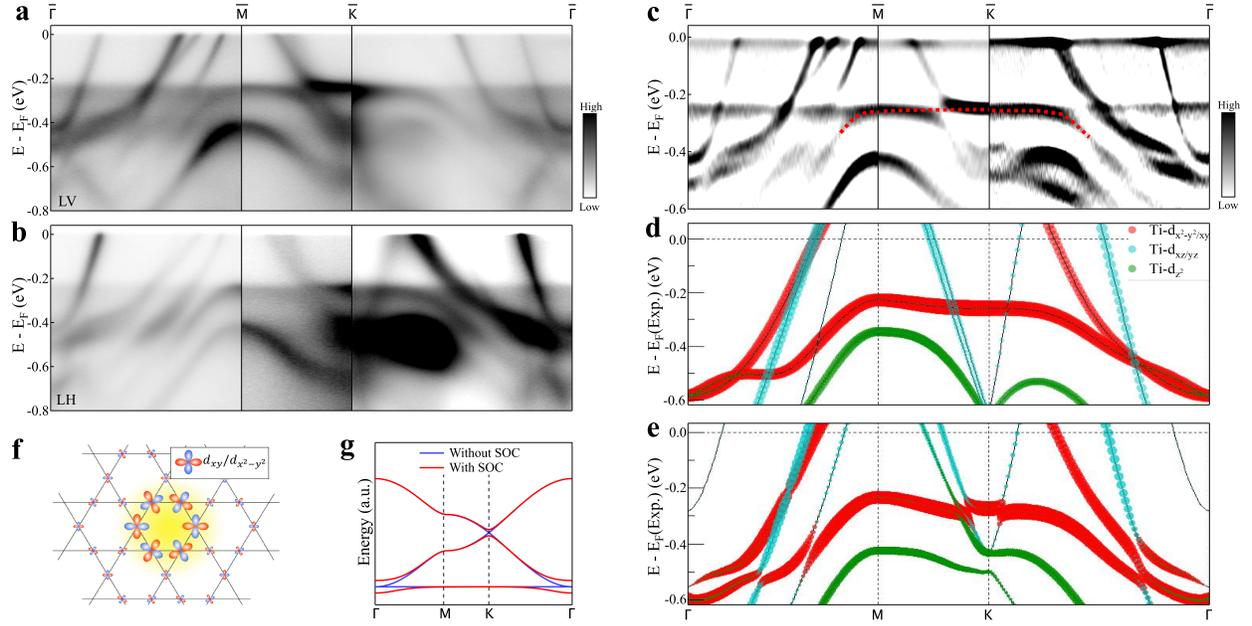}
\end{center}
\caption{\textbf{Direct observation of flat band in CsTi$_3$Bi$_5$.} \textbf{a,b} Detailed band structures measured  along the $\bar{\Gamma}$-$\bar{M}$-$\bar{K}$-$\bar{\Gamma}$ high symmetry directions under two different polarization geometries, LV (\textbf{a}) and LH (\textbf{b}). \textbf{c}  Second derivative image with respect to energy obtained from \textbf{a,b} in order to resolve the band structures more clearly. The red dashed line highlights the observed flat band. \textbf{d} The corresponding calculated band structure along the $\Gamma$-M-K-$\Gamma$ high-symmetry directions  without considering SOC. Different colors represent different orbital components of Ti$_{3d}$. \textbf{e} Same as \textbf{d} but considering SOC. \textbf{f} Orbital textures of the effective Wannier states giving rise to the flat bands with d$_{xy}$/d$_{x^2-y^2}$ orbitals. \textbf{g} Tight-binding band structures of kagome lattice with (red lines) and without (blue lines) SOC. Inclusion of the spin orbit coupling gaps both the Dirac crossing at K and the quadratic touching between the flat band and the Dirac band around $\Gamma$.
}
\end{figure*}

\begin{figure*}[tbp]
\includegraphics[width=1.0\columnwidth,angle=0]{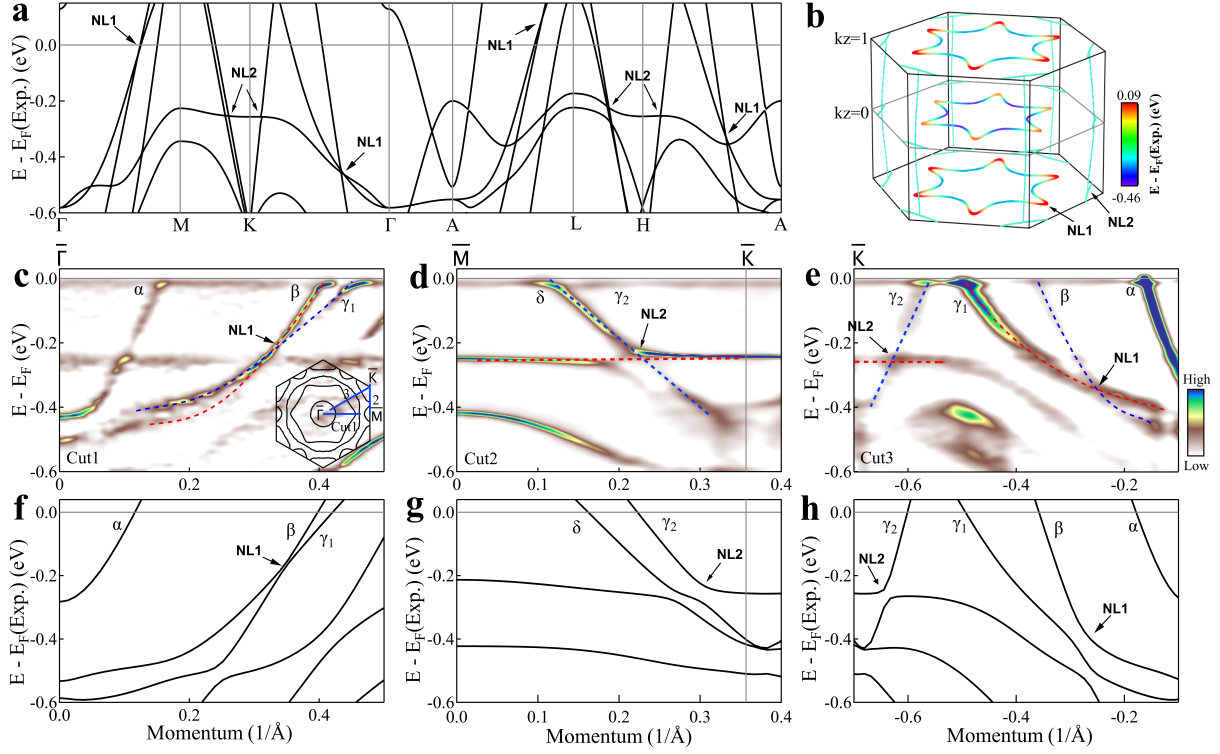}
\begin{center}
\caption{\textbf{Calculated and observed Dirac nodal points in CsTi$_3$Bi$_5$}.  \textbf{a} Calculated band structures along high-symmetry directions without considering SOC. Two sets of Dirac nodal points (NL1 and NL2) are identified as marked by arrows. \textbf{b} Formation of the Dirac nodal lines in three dimensional momentum space. The NL1 Dirac points form hexagonal Dirac nodal loops around $\Gamma$ in the k$_z$=0 plane and A in the k$_z$=1 plane. The NL2 Dirac points form triangular Dirac nodal loops around K in the k$_z$=0 plane and H in the k$_z$=1 plane as well as six nodal lines along the k$_z$ direction. The detailed distribution of the NL1 and NL2 Dirac nodal lines in three-dimensional momentum space is shown in Fig. S2 in Supplementary Materials.  \textbf{c-e} Band structures measured along $\bar{\Gamma}$-$\bar{M}$ (Cut1), $\bar{M}$-$\bar{K}$ (Cut2) and $\bar{K}$-$\bar{\Gamma}$ (Cut3), respectively. The location of the momentum cuts is shown in the inset of \textbf{c} by the blue lines. These images are obtained by taking second derivative curvature of the original data. The blue and red dashed lines are guide lines of the two crossing bands. \textbf{f-h} The corresponding calculated band structures with SOC. The presence of the Dirac points is marked by arrows.
}
\end{center}
\end{figure*}

\begin{figure*}[tbp]
\begin{center}
\includegraphics[width=1.0\columnwidth,angle=0]{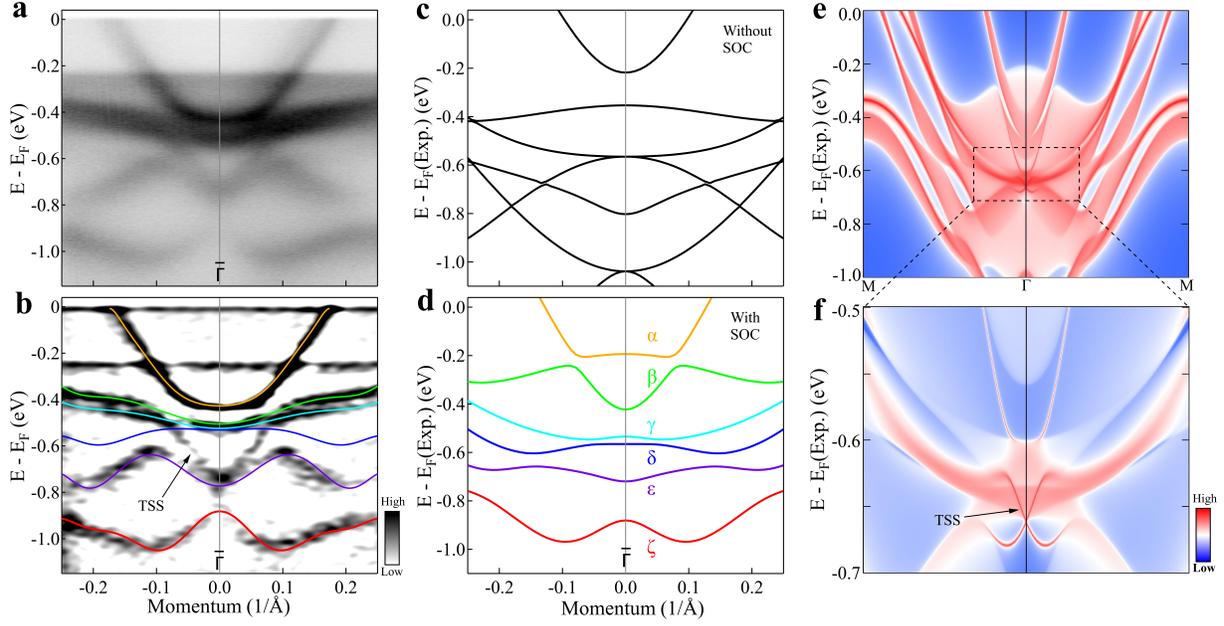}
\end{center}
\caption{\textbf{Observation of topological surface states in CsTi$_3$Bi$_5$}. \textbf{a} Band structure measured around $\bar{\Gamma}$ along the $\bar{M}$-$\bar{\Gamma}$-$\bar{M}$ direction under LV polarization geometry. \textbf{b} Second derivative image of \textbf{a}. The observed six bands are highlighted by different colored lines. The observed topological surface state (TSS) is marked by an arrow. \textbf{c-d} The corresponding calculated band structures without (\textbf{c}) and with (\textbf{d}) SOC. The topological surface state emerges between the $\delta$ and $\epsilon$ bands which arise from band inversion due to SOC. \textbf{e} Calculated surface spectral function along $ \bar{M} $-$ \bar{\Gamma} $-$ \bar{M} $ paths projected on the (001) plane for the Bi-terminated CsTi$_3$Bi$_5$. \textbf{f} Enlarged view of the topological surface states (TSS) in \textbf{e}.
}
\end{figure*}

\end{document}